# Reflector Antennas Characterization and Diagnostics using a Single Set of Far Field Phaseless Data and Crosswords-like Processing

R. Palmeri, G. M. Battaglia, A. F. Morabito, and T. Isernia, *Senior Member, IEEE*

*Abstract*—We introduce and discuss a new approach to the phase retrieval of fields radiated by continuous aperture sources having a circular support, which is of interest in many applications including the detection of shape deformations on reflector antennas. The approach is based on a decomposition of the actual 2-D problem into a number of 1-D phase retrieval problems along diameters and concentric rings of the visible part of the spectrum. In particular, the 1-D problems are effectively solved by using the spectral factorization method, while discrimination arguments at the crossing points allows to complete the retrieval of the 2-D complex field. The proposed procedure, which just requires a single set of far field amplitudes, takes advantage from up to now unexplored field properties and it is assessed in terms of reflector aperture fields.

## I. Introduction

IN THE framework of reflector antennas, deviations of the surface from the ideal reflector surface caused by gravity, wind, load, temperature play an important role as they imply a degradation of the antenna efficiency [1],[2]. In fact, as discussed in [1], the root mean square error (RMSE) deviation of the reflector surface from the ideal one should be less than $\lambda/15$ in order to obtain more than 50% aperture efficiency ($\lambda$ denoting the operating wavelength). Therefore, the challenge is to retrieve the shape in an accurate fashion, so that possible countermeasures can be adopted. In this respect, it is worth noting that Rahmat-Samii introduced in [2] a simple and widely adopted model relating the reflector shape deformation to the phase of the aperture field. More recently, a deformation-amplitude relationship has also been introduced in [3]. In both cases it is argued that the knowledge of the aperture field allows to provide the desired information on deformation.

Several approaches have been proposed for reflector surface diagnostics. In short, one can find strategies based on the processing of both amplitude and phase of the radiated far field [4]-[8] as well as strategies exploiting just the far field power pattern (see for instance [9]-[14]). Since collecting phase measurements can be very difficult or excessively expensive with increasing frequencies [15]-[17], the second kind of strategy is indeed of interest. Accordingly, the problem of the surface shape detection in reflector antennas can be conveniently linked to a Phase Retrieval (PR) problem [10],[13].

By considering an unknown complex function $f(\underline{x})$ and an operator $\mathcal{T}$ such that $F(\underline{k}) = \mathcal{T}[f(\underline{x})] = |F(\underline{k})|e^{j\angle F(\underline{k})}$, the PR problem deals with the determination of the complex quantity $F(\underline{k})$ starting from the knowledge of $|F(\underline{k})|$ plus some additional information [18].

A very popular approach for such a kind of problems has been given by Misell in [9] in the context of electron microscopy, and later adapted to reflector antennas [10]. As essentially all PR methods available in literature for antenna or other devices characterization (but for [17],[19]), this approach requires a given diversity in collecting data (e.g., two or more probes, two or more measurement surfaces, defocus conditions, or the like [20],[21]). While being conceptually simple and fast, this procedure may converge to a so called 'false' solution [22] if the initial estimation of the aperture distribution is far from the ground truth. A smart way to overcome this issue (at least from a practical point of view) is the so-called 'hybrid input-output' empirical strategy [23], which is widely adopted [24].

Another benchmark and well-assessed approach, based on global optimization, is proposed in [25]. However, since the computational complexity of global optimization-based procedures is expected to exponentially grow with the number of unknowns [26], it can prevent the actual attainment of the global optimum in case of very large sources.

Finally, different approaches to PR have been recently introduced by relying on the 'PhaseLift' strategy (see [27],[28] and references therein). Unfortunately, the latter has two important drawbacks, i.e., the computational complexity (which grows very rapidly with the number of unknowns) and the capability of determining only one solution of the problem (while in many cases, including 1-D PR, the problem can admit multiple solutions) [29].

This work was supported in part by the Italian Ministry of University and Research under the PRIN research project "CYBER-PHYSICAL ELECTROMAGNETIC VISION: Context-Aware Electromagnetic Sensing and Smart Reaction", prot. 2017HZJXSZ.

R. Palmeri is with Institute for the Electromagnetic Sensing of the Environment of the National Council of Research (IREA-CNR), 80124 Napoli, Italy. G. M. Battaglia, A. F. Morabito, and T. Isernia are with Department of Information Engineering, Infrastructures, and Sustainable Energy, Università di Reggio Calabria, 89122 Reggio Calabria, Italy.

Authors are also with CNIT (Consorzio Nazionale Interuniversitario per le Telecomunicazioni), 43124 Parma, Italy.

Corresponding author: Tommaso Isernia (tommaso.isernia@unirc.it).

As a contribution towards effective solution strategies for such a problem, we present in the following a novel approach to PR that avoids the exploitation of global-optimization algorithms. In particular, by exploiting only one set of far field amplitude measurements (in combination with the knowledge of the source support and some other minimal information), it allows determining the globally-optimal solution.

The proposed approach relies on three basic bricks.

The first, and more obvious one, is the aperture antenna theory, which allows for simple relationships amongst the aperture field and the far field, from one side, and the aperture field and the reflector deformations, from the other side.

The second brick, related indeed to the previous one, is given by the expansions of the aperture and far field in terms of the so called Orbital Angular Momentum (OAM) modes, where the far field is given by suitable Hankel transforms of the aperture field modes [30],[31].

The third (and most clever) brick is a recent approach to PR problems introduced by the authors for the case of 2-D discrete signals (with particular attention to array antennas and hence periodic spectra) [29],[32]. In the latter, an analogy between filling the rows and columns of the spectrum matrix and completing a 'crosswords' scheme is introduced and discussed. In particular, the approach in [29],[32] first casts the 2-D PR as a combination of 1-D PR problems. Then, it finds in a deterministic fashion all the admissible solutions of each 1-D problem by means of the Spectral Factorization (SF) method [33],[34]. Finally, enforcing congruence and discrimination criteria amongst them, the approach determines the correct field behavior along the different (horizontal, vertical, and diagonal) strips covering the 2-D scenario.

As in [29],[32], the approach presented herein also exploits the strategy of decomposing the 2-D PR into a number of 1-D problems and solving these latter through the SF method. However, differently from [29],[32], we solve herein 1-D problems along circles and diameters (rather than horizontal, vertical, and diagonal strips) over the spectral plane. As discussed in the following (and briefly anticipated in [35]), such a choice automatically allows a considerable reduction of the computational burden associated to the inherent combinatorial problem. Moreover, it also allows taking full advantage from effective representations based both on OAM field modes [30] and singular value decompositions (SVD) [36] of the fields corresponding to each mode. These representations lead to a simple initialization of the PR procedure as well as a considerable reduction of the number of cases to be considered when checking congruence amongst fields on diameters and circles, which is achieved by exploiting a relevant (and up to now overlooked) property of the fields along circles of the spectral plane.

Once the field has been determined in a number of circles and diameters of the spectral plane, the overall scheme is completed by solving a fitting problem that takes into account both the starting data (i.e., the square amplitude of the far field) and the complex field previously retrieved in a subset of the sampling points. In so doing, the presence of a quadratic and positive-definite term allows (along the guidelines of [22],[37]) avoiding the occurrence of false solutions.

Notably, the capability to complete the overall PR procedure by exploiting a single set of measurements (i.e., no diversity is needed) is indeed a relevant asset, since it allows a great simplification in the measurement procedure and acquisition time [17].

The method is herein presented in case of generic planar continuous sources contained having a finite circular support and exhibiting no particular symmetries. By virtue of its generality, it applies to whatever related PR problem, so that similar tools may be applied to synthesis (rather than recovery) problems.

The remainder of the paper is as follows. In Section II, we briefly recall some useful mathematical tools of interest for aperture antennas. Then, convenient representations for the aperture field and the corresponding spectrum are given in Section III. In particular, it is argued that the aperture field's spectrum can be represented as a trigonometric polynomial along any circle and diameter of the spectral domain, and some up to now overlooked properties of this kind of polynomials are given. Then, the rationale of the proposed strategy is presented in Section IV, while the procedure arising from the joint consideration of the basic approach and of the introduced properties is given in Section V. Finally, Section VI is devoted to an application of the given concepts to the diagnostics of deformations of reflector antennas. Conclusions follow.

## II. SOME USEFUL EXPANSIONS AND FORMULATION OF THE PROBLEM

By the sake of simplicity, let us consider the case where the aperture field is purely polarized along the $x$ (or $y$) direction. In such a case, it is simple to see from the aperture antenna theory [38] that the knowledge of the square amplitude of the far field components $E_{\infty_\theta}(\theta,\phi)$ or $E_{\infty_\phi}(\theta,\phi)$, $\theta$ and $\phi$ respectively denoting the antenna elevation and azimuth angles, is equivalent to the knowledge of the square amplitude of the Fourier transform of the aperture field components $E_x(x,y)$ [or $E_y(x,y)$] in the visible part of the spectrum, $x$ and $y$ denoting the coordinates spanning the aperture plane. Hence, assuming we can measure either $\left|E_{\infty_\theta}\right|^2$ or $\left|E_{\infty_\phi}\right|^2$, we can restrain our attention to the 2-D Fourier-transform relationship amongst the (scalar) aperture field, named $f(x,y)$ in the following, and the corresponding far field.

Then, let us consider a continuous aperture field $f$ having a circular support of radius $a$, which is of interest in reflector antenna diagnostics and in many other cases[1] [2]. Notably, by denoting with $\rho'$ and $\phi'$ the radial and azimuth coordinates spanning the aperture, respectively, $f$ can be expanded in a multipole series as (see [30] for more details):

---

[1] Obviously, any finite-dimensional planar source can be delimited by some circle having a sufficiently large radius.

$$f(\rho', \phi') = \sum_{\ell=-\infty}^{+\infty} f_\ell(\rho') e^{j\ell\phi'} \quad (1)$$

where:

$$f_\ell(\rho') = \frac{1}{2\pi} \int_0^{2\pi} f(\rho', \phi') e^{-j\ell\phi'} d\phi' \quad (2)$$

Then, by denoting with $k'$ and $\phi$ the radial and azimuth coordinates spanning the spectral domain, the Fourier transform of the source (1) is equal to:

$$F(k', \phi) = \frac{1}{2\pi} \int_0^{2\pi} \int_0^\infty f(\rho', \phi') e^{-jk'\rho'\cos(\phi'-\phi)} \rho' d\rho' d\phi' \quad (3)$$

which can also be expanded in a multipole series as:

$$F(k', \phi) = \sum_{\ell=-\infty}^{+\infty} F_\ell(k') e^{j\ell\phi} \quad (4)$$

where:

$$F_\ell(k') = \frac{1}{2\pi} \int_0^{2\pi} F(k', \phi) e^{-j\ell\phi} d\phi \quad (5)$$

On the basis of the (polarization) assumptions above, let us now consider the problem of reconstructing the 2-D source $f(\rho', \phi')$ from the knowledge (e.g., measurements) of the square-amplitude distribution of its Fourier transform in the visible part of the spectrum, i.e., $k' \leq \beta$ ($\beta$ being the wavenumber).

By denoting with $\mathcal{B}[\cdot]$ the operator performing the square amplitude operation, i.e.:

$$\mathcal{B}[F(k', \phi)] = F(k', \phi) F^*(k', \phi) = |F(k', \phi)|^2 \quad (6)$$

the 2-D PR problem at hand can be stated as the determination of the 'correct' $F$ distribution (say $\hat{F}$) such that:

$$\mathcal{B}[\hat{F}(k', \phi)] = M^2(k', \phi) \quad (7)$$

where $M^2(k', \phi)$ denotes the measured square-amplitude far-field distribution[2].

It is worth to underline that the problem of retrieving the phase of $\hat{F}$ (in the visible part of the spectrum) and retrieving instead $f(\rho', \phi')$ are not exactly the same, as the whole spectrum (and not just its visible part) would be required in order to safely recover the source distribution. Also, let us explicitly note that in case of noisy data, the measured square-amplitude distribution may not belong to the range of $\mathcal{B}$ and hence, even if uniqueness issues (see below) are neglected, the problem is still ill-posed [18]. In fact, one generally looks for some best fitting amongst the two quantities in (7), rather than to an exact equality [18].

### III. A CONVENIENT REPRESENTATION FOR FIELDS AND SOURCES, AND SOME USEFUL PROPERTIES

#### A. Representing field and source via OAM modes and SVD

To give an accurate representation of the field, we can exploit the SVD of the radiation operator relating the source $f_\ell(\rho')$ to the field $F_\ell(k')$.

In fact, if (1) is substituted into (3), and then (4) is used, the contribution to the radiated field given by (5) can be written as:

$$F_\ell(k') = \int_0^a f_\ell(\rho') J_\ell(k'\rho') \rho' d\rho' = H_\ell\{f_\ell(\rho')\} \quad (8)$$

wherein, as reported in [30], $a$ is the radius of the circle enclosing the planar source at hand, $J_\ell(\cdot)$ is the $\ell$-th order Bessel function of the first kind, and the last expression denotes the $\ell$-order Hankel transform of $f_\ell(\rho')$ (which is supposed to be zero for $\rho' > a$) [31].

Then, equation (8) can be written in an operator form as:

$$F_\ell = A_\ell f_\ell \quad (9)$$

$A_\ell$ being a compact notation for the corresponding (radiation) operator.

We can now perform the SVD of $A_\ell$, i.e., $\{v_{\ell,n}, \sigma_{\ell,n}, u_{\ell,n}\}$ such that:

$$A_\ell v_{\ell,n} = \sigma_{\ell,n} u_{\ell,n} \quad (10a)$$
$$A_\ell^+ u_{\ell,n} = \sigma_{\ell,n} v_{\ell,n} \quad (10b)$$

$\sigma_{\ell,n}$, $v_{\ell,n}$, $u_{\ell,n}$ denoting the $n$-th singular value, right-hand singular functions, and left-hand singular functions associated to the $\ell$-th order, respectively, while $A_\ell^+$ is the adjoint operator of $A_\ell$. A detailed description on how to compute the functions and scalars at hand is given in [30].

Since the singular functions $v_{\ell,n}$ and $u_{\ell,n}$ are orthonormal in the space of sources and fields [30],[36], respectively, they can be used as representation bases in the corresponding domains. Therefore, we can represent the different field components as follows:

$$F_\ell(k') = \sum_{n=1}^\infty b_{\ell,n} u_{\ell,n}(k') \quad (11)$$

and, by substituting (11) into (4), we finally achieve:

$$F(k', \phi) = \sum_{\ell=-\infty}^{+\infty} \sum_{n=1}^\infty b_{\ell,n} u_{\ell,n}(k') e^{j\ell\phi} \quad (12)$$

Then, assuming the source is not superdirective, expansion (12) can be conveniently truncated by following the rules reported in [30]. By so doing, one achieves:

$$F(k', \phi) = \sum_{\ell=-\beta a}^{\beta a} \sum_{n=1}^{N_\ell} b_{\ell,n} u_{\ell,n}(k') e^{j\ell\phi} \quad (13)$$

where $N_0 = \frac{2a}{\lambda}$ and the generic value of $N_\ell$ is given by:

$$N_\ell = N_0 - \frac{|\ell|}{\pi} \quad (14)$$

Notably, starting from (13), a convenient representation for the aperture field, corresponding to a regularized inversion from the (visible part of the) spectrum to the source, is given by:

---

[2] Note that, by virtue of the bandlimitedness of the spectrum [39], the function $M^2(k', \phi)$ can be conveniently acquired through proper sampling and interpolation operations.

$$f(\rho, \phi') = \sum_{\ell=-\beta a}^{\beta a} \sum_{n=1}^{N_\ell} \alpha_{\ell,n} v_{\ell,n}(\rho) e^{j\ell\phi'} \quad (15)$$

with $b_{\ell,n} = \sigma_{\ell,n} \alpha_{\ell,n}$.

*B. Trigonometric polynomial representation of the spectrum along circles of the spectral domain, and their properties with increasing radii*

To achieve field and source representations even more convenient than (13) and (15), advantage can be taken of the fact that the singular functions $u_{\ell,n}(k')$ exhibit a $|\ell|$-th order zero for $k' = 0$. Hence, for $k' = 0$ only the functions $u_{0,n}(k')$ contribute to the field. More generally, the smaller the value of $k'$ the smaller the value of $\ell$ up to which functions $u_{\ell,n}(k')$ play a significant role in the field generation [30].

Accordingly, in (13) the external summation can be further truncated to the interval $[-H, H]$, where $H$ depends on both $a$ and $k'$.

For all the above, (13) can be written as:

$$F(k, \phi) = \sum_{\ell=-H(k',a)}^{H(k',a)} \sum_{n=1}^{N_\ell} b_{\ell,n} u_{\ell,n}(k) e^{j\ell\phi} \quad (16a)$$

Therefore, when considering a ring of radius $k' = \overline{k}$, the field can be conveniently written in terms of a trigonometric polynomial whose order depends indeed on $\overline{k}$ (i.e., the lower $\overline{k}$, the lower the order of the polynomial). As a matter of fact, along a ring of radius $\overline{k}$ one can use:

$$F(\overline{k}, \phi) = \sum_{\ell=-H(\overline{k},a)}^{H(\overline{k},a)} C_\ell(\overline{k}) e^{j\ell\phi} \quad (16b)$$

As far as the actual choice of $H$ is concerned, it could be deduced from the properties of the singular functions. However, a simpler yet accurate estimation of the (minimum) value of $H$ to be used in (16b) can be given as follows.

First note that, by virtue of the source size, the minimal sampling period in the spatial frequency domain is equal to $1/2a$. As a consequence, the sampling period along any line in the (angular frequency) $k'$ domain is equal to $\pi/a$. Hence, along a circumference having a radius $k' = \overline{k}$ one needs a number of samples equal to $(2\pi\overline{k})/(\pi/a) = 2\overline{k}a$.

Finally, by performing a trigonometric interpolation of these samples, one can transform the arising sampling series based on the Dirichlet kernel into a truncated Fourier series [40], finding that $H$ must be greater than or equal to $\overline{k}a$.

The choice $H = \overline{k}a$, which we assume in the following, nicely agrees with (13)-(16) when $\overline{k} = \beta$, i.e., along the maximum circle of the visible space. In fact, $H(\beta, a) = \beta a$.

Notably, by obvious derivations, the square amplitude distribution of the fields along rings can also be expressed as trigonometric polynomials whose order is exactly twice the one of the corresponding fields. Such a circumstance also suggests a further (practical) rule for a convenient choice of $H$ along any circle in the spectral domain. In fact, one can estimate the value of $2H$ directly from the square amplitude distribution available along the circle at hand.

Expressions (16) for the field in terms of the singular functions (or, more simply, in terms of trigonometric polynomials) can be conveniently used in the proposed procedure. To this end, let us note that trigonometric polynomials of the kind (16b) can also be seen as the restriction to the unitary circle of an expression of the following kind:

$$F(\overline{k}, \phi) = \sum_{\ell=-H}^{H} C_\ell(\overline{k}) z^\ell \quad (17)$$

with $z = e^{j\phi}$. It is also well known [41] that the complex zeroes of (17) determine the actual behavior of the field [33].

In this respect, it is useful to get an understanding of the (number and) locations of these zeroes with increasing values of $k'$, which corresponds to increasing values of $H$.

One can easily understand that such an order is nothing but zero when $k' = 0$. Then, when $k'$ gently moves from the origin to larger values, the order of the polynomial also progressively increases, so that new zeroes come into play in the $z$-plane. Notably (see Appendix A) if one considers two nearby concentric circles with different radii such that the required value of $H$ increases by one, and no zero occurs in the corresponding annulus of the spectrum, the two additional zeroes must necessarily be located one inside and one outside the unitary circle in the $z$-plane[3]. Then, as long as the function $F(k', \phi)$ does not exhibit any zero in the circle $k' < \overline{k}$, the trigonometric polynomial (17) exactly has half of the zeroes inside the unitary circle and the other half outside[4].

*C. Trigonometric polynomial representation of the spectrum along diameters of the spectral domain*

It is also interesting to look for convenient field representation for fixed values of $\phi$, i.e., along diameters of the circle corresponding to the visible region of the spectral plane. If we assume that the spectrum is negligible in its invisible part, and that it is sufficiently small on its border, (which is usually the case with reflector antennas) the behavior of the spectrum can be accurately reconstructed from its Nyquist samples along the diameter at hand. Then, as any diameter of the visible space is shorter than the maximum circle by a factor $\pi$, one finds that the number of required samples is equal to $2\beta a/\pi = 4a/\lambda$, and the same statement holds true if we re-normalize the coordinate along the diameter by using $k'' = k'/2$, so that $-\pi \leq k'' \leq \pi$. Finally, for any of the above diameters, identified by $\phi = \overline{\phi}$, we can again use the correspondence amongst Dirichlet sampling series and truncated Fourier series [40] to come to the following expression:

---

[3] Actually, they will be close to the origin and close to infinity, respectively, when beginning coming into play.

[4] If $F(k', \phi)$ exhibits instead one or more zeroes within the circle $k' < \overline{k}$, one can eventually track the evolutions of the zeroes in the complex plane, thus determining (for any fixed value of $\overline{k}$) how many zeroes of the corresponding trigonometric polynomial are inside or outside the unitary circle of the $z$ plane.

$$F(k', \overline{\phi}) = \sum_{\ell=-2a/\lambda}^{2a/\lambda} \hat{C}_\ell(\overline{\phi}) z'^{\ell} \qquad (18)$$

with $z' = e^{\frac{jk'}{2}} = e^{jk''}$ [5].

A more detailed explanation of (18) can be found in [42] where it is also argued that it may be convenient to slightly extend the summation in case of sources which are not so large with respect to $\lambda$. Roughly speaking, one can also argue that (18) is nothing but the field radiated by the original source as collapsed (according to the definition in [29],[43]) on the $\phi = \overline{\phi}$ line, and it is known [42] that the field radiated by a linear non-superdirective source can be seen as the field radiated by an equivalent 'virtual' array with a spacing equal to (or slightly smaller) than $\lambda/2$. In fact, (18) can also be seen as an array factor where the present $k''$ replaces the usual spectral variable '$u$'.

Again, the (complex) zeroes of (18) will determine the actual behavior of the field [33], and of course the square amplitude distribution of the spectrum will have a similar expansion (with $4a/\lambda$ replacing $2a/\lambda$). Finally, since the source is supposed to be circular, the summation indices do not vary with the chosen line, i.e., with the value of $\overline{\phi}$. However, different indices could be conveniently used for different $\phi$ values in other cases, e.g., elliptical planar sources.

## IV. RATIONALE OF THE PROPOSED PHASE RETRIEVAL STRATEGY: A CONCENTRIC CROSSWORDS-LIKE SCHEME

The rationale, scope, and capabilities of the proposed approach to 2-D PR are intimately related to well-known results regarding the uniqueness (or lack of uniqueness) for the considered problem in both the 1-D and 2-D cases.

As a matter of fact, 1-D PR problems do not admit a unique solution. In fact, besides the 'trivial ambiguities' discussed below, the far-field square amplitude can be written (but for a constant) as the product of factors of the kind[6] $e^{j\hat{u}} - z_i$, $e^{j\hat{u}} - \frac{1}{z_i^*}$, and any 'flipping' amongst each couple corresponds to a different solution[7] [33]. On the other side, by means of a well-defined procedure based on extracting (and pairing) the roots of the field expression, one is able to determine all the possible solutions of the 1-D PR problems at hand. Therefore, for each 1-D PR problem we have a collection of possible solutions for the pertaining part of the spectrum.

As 2-D polynomials are not factorable (but for a zero-measure set of cases), ambiguities due to the spectrum factorization do not generally occur in 2-D PR problems [17]. Accordingly, the solution of a number of 1-D PR problems and their correct intersection according to congruence and discrimination arguments allow achieving the actual solution, the only residual problem being the so called 'trivial ambiguities'. These latter consist in:

i) a constant phase on the far field/aperture source;
ii) a linear phase on the far field (just possible along diameters), corresponding to a translation of the source;
iii) a conjugation of the complex far field, corresponding to a reversal plus a conjugation of the source (i.e., of the aperture field in the overall problem considered herein);
iv) any combination of the above.

In fact, each of the above situations results in the same identical square-amplitude field distribution.

The first two ambiguities can be successfully faced by fixing a reference phase and supposing the support of the source to be known, respectively. The third kind of ambiguity keeps however there, and hence (even neglecting the zero-measure set of cases where the 2-D spectrum can be further factored) some additional a-priori information is needed in order to get a theoretically unique solution also in the 2-D case.

By assuming that such additional information is available, or when looking for all the different solutions in case of non-uniqueness, we still need a computational procedure able to get the ground truth (in case of uniqueness) or all the possible solutions (in case uniqueness does not hold true). As a matter of fact, because of false solutions [22], or computational issues [17],[21],[37],[44], this is still an open problem in the literature, unless a number of independent measurements of the radiated field substantially larger than the ones required for the theoretical uniqueness is used. This is actually the reason why in antenna diagnostics two different sets of (sufficiently different) phaseless measurements [20],[21],[45]-[52] are usually assumed (but for [17],[19],[29],[32]).

Our proposed solution procedure, originally introduced for the case of array antennas in [29],[32], is inspired from the solution of crosswords puzzles. These latter, which are quite diffused in every culture whose language adopts alphabets rather than ideograms, are indeed puzzles where vertical and horizontal words, subject to some (possibly ambiguous) definitions, have to correctly intersect. In fact, by virtue of bandlimitedness, the far-field matrix to be retrieved can be seen as a crossword scheme to be solved, so that the 2-D PR problem has been recently tackled as a collection of auxiliary 1-D PR problems along rows and columns.

Since we are dealing with circularly-supported aperture sources, and by virtue of the fact that the adopted choice brings a number of advantages, herein we will consider instead 1-D PR problems along rings and diameters of the spectrum. For this reason, we refer to the present approach as to a 'concentric' crosswords-like scheme.

The approach overcomes relevant issues associated to our previous approach in [29],[32]. In fact, in the latter, a horizontal, a vertical, and a diagonal cut of the far field were needed in order to have three intersections and initialize the solution of the puzzle, resulting in a cumbersome combinatorial problem[8]. Herein, the 'concentric' scheme allows instead starting with just one ring and one diameter and hence considerably reducing the computational burden[9]. Moreover, and more important, as discussed in Section V.B, further relevant advantages do occur.

---

[5] Note such a choice implies that spanning a diameter corresponds to spanning the entire unitary circle in the $z$ plane.
[6] $\hat{u} \equiv k'/2$ along diameters, whereas $\hat{u} \equiv \phi$ along rings.
[7] Unless the two roots of the couple have unitary amplitude.

[8] In fact, the first two intersections just normalize the phase along the second and third strip, and the third intersection allows for discrimination [29],[32].
[9] In fact, they have two intersection points, the second one already allowing to discard incongruent solutions amongst the chosen diameter and ring.

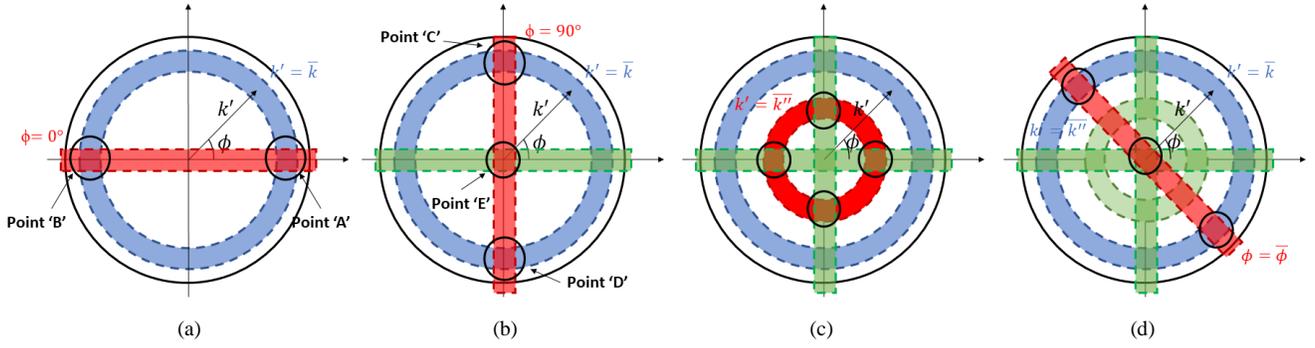

(a) (b) (c) (d)

Fig. 1. 'Concentric crosswords' scheme of the proposed PR strategy. The continuous black circle indicates the visible space limit. (a) Intersections between the $\phi = 0$ diameter and the $k' = \overline{k}$ ring, the complex field being known in point 'A' (which is used for phase normalization) and in point 'B' (which allows for discriminating unsuitable solutions). (b) Intersection amongst the $\phi = 90°$ diameter, the $k' = \overline{k}$ ring, and the $\phi = 0°$ diameter. One of the three intersections can be used for adjusting the phase constant, and the other two for discriminating unsuitable 1-D solutions along the vertical diameter. (c) Intersections amongst the $k' = \overline{k''}$ ring and the two diameters respectively given b $\phi = 0°$ and $\phi = 90°$ (all intersection points being known from previous intersections). (d) Intersections amongst the $\phi = \overline{\phi}$ diameter and the already considered rings.

## V. THE PROPOSED PROCEDURE

In the proposed procedure, consisting of two steps, one has to solve a sequence of 1-D problems and then use congruence arguments at the intersection points until the entire far field matrix is retrieved. As already noticed, a substantial speed-up is however possible once solutions along a sufficient number of rings and diameters have been identified.

In the following, the two steps of the overall procedure are described in the subsections (V.A) and (V.D),(V.E), respectively, whereas subsections (V.B) and (V.C) respectively discuss its drawbacks and the corresponding countermeasures and improvements allowed by the properties discussed in subsection III.B.

### V.A. Step 1: 2-D Phase Retrieval via 1-D Spectral Factorization and discrimination arguments

Let us define with $\hat{u}$ the generic variable spanning the 1-D observation domain at hand, i.e.:

$$F(\hat{u}) = \begin{cases} F(k'/2) & for\ \phi = \overline{\phi}\ \ (i.e., a\ diameter) \\ F(\phi) & for\ k' = \overline{k}\ \ (i.e., a\ ring) \end{cases} \quad (19)$$

Then, by using (17) or (18), and by obvious derivations [33], the square-amplitude distribution of (19) can be expressed in terms of auxiliary coefficients $D_p$ such that:

$$|F(\hat{u})|^2 = S(\hat{u}) = \sum_{p=-2P}^{2P} D_p e^{jp\hat{u}} \quad (20)$$

where, according to contents of Section III, $P = H = \overline{k}a$ when $\hat{u} = \phi$ (i.e., for a ring), and $P = H = 2a/\lambda$ when $\hat{u} = k'/2$ (i.e., for a diameter). Also note that because of the fact that the left-hand member is a real quantity, $\{D\}$ is a Hermitian sequence of $4P + 1$ complex coefficients, i.e., $D_p = D^*_{-p}$, $p = 0, 1, …, 2P$.

Then, by considering the measured square-amplitude data, i.e., $M^2$ as sampled in a grid of points $\mathbf{u} = [\hat{u}_1, …, \hat{u}_N]$, where $N$ is the number of measurements taken according to [39], the Hermitian sequence $\{D_p\}$ can be identified by enforcing that:

$$\sum_{p=-2P}^{2P} D_p e^{jp\mathbf{u}} = M^2(\mathbf{u}) \quad (21)$$

Once the coefficients $\{D\}$ have been determined, the SF technique developed in [33] can be applied in order to get the multiplicity of solutions for the pertaining 1-D cut (ring) of the power pattern. Then, a 'concentric crosswords' scheme can be applied as described in the following with the help of Fig. 1.

In order to better understand the ratio and the difficulties of the problem, and how possibly get rid of it, let us initially consider the (horizontal) diameter and a generic ring, as depicted in Fig. 1(a). Also, let us suppose by the sake of simplicity that we also know the phase of the fields at the two intersection points[10], and that our measurements are noiseless. Both the simplifying assumptions will be removed at a later stage.

Both along the considered diameter and the considered ring, a number of possible solutions are available through the SF method. In order to identify the admissible ones, point 'A' can be used as a normalization point for the field phase, while point 'B' can be used to discard a number of unsuitable (potential) solutions. In fact, the only admissible solutions are the couples of solutions (along the considered diameter and the considered ring) having the same exact (known) phase value at the point B. Hence, one can hopefully identify the correct complex field distribution along the initial diameter and the initial ring.

Once this task has been performed, it is possible to consider another cut, e.g., the vertical diameter [see Fig. 1(b)], and to solve (21) along it. Notably, one has now three points where field amplitude and phase knowledge is available, as point 'E' is known from the knowledge of the field along the horizontal diameter while points 'C' and 'D' are available from the knowledge of the field along the ring. Then, one of these points can be used to normalize the field phase along the vertical line at hand, and the other two points will allow for pruning the set of possibilities along the line and get the correct field behavior along this second diameter.

---

[10] Of course, as we can fix the phase to zero at the first intersection point, we just need the phase shift at the second point.

As exemplified through Fig. 1, one can now iterate the procedure by considering additional rings [see Fig. 1(c)] and additional diameters [see Fig. 1(d)]. Notably, which is a relevant asset of the procedure, identification of the correct field behavior along diameters and rings becomes easier and easier when proceeding with new rings and diameters. In fact, one has at her/his disposal an increasing number of discrimination points for each 1-D PR problem.

As a general warning, one should avoid using as intersection or discrimination points the ones where the field is zero (or nearly zero). In fact, in those points phase simply makes no sense (or it may be affected by a large error in case of noisy data).

Obviously, as in an actual crosswords puzzle solution, it makes sense starting from the 'easiest' part of the spectrum. In particular, as far as diameters are concerned, it makes sense to start from field cuts where many zeroes are present, as they allow dealing with a reduced number of multiple solutions in the corresponding 1-D SF problem [33], and hence with a reduced computational burden. As far as rings are concerned, it is intuitively convenient to start from circles having a small radius, as they correspond to 'words' to be identified which are shorter and hence hopefully simpler. A detailed discussion of this point is deferred to subsection V-C below.

*V.B. Limitations of the basic approach*

As its companion method presented in [29],[32], the proposed approach is affected from two main (related) limitations.

First, the presence of noise on data implies that one has to relax the requirements at the discrimination points. In fact, some tolerance has to be guaranteed when checking congruence in order to avoid discarding potential solutions along the different domains. As a consequence, it may become harder and harder to come to a unique solution (if any).

Second, the larger the considered diameter or the considered circle, the higher the order of the involved polynomials. As a consequence, one will have very many candidate solutions along the line (or the ring) at hand, so that many of them could be anyway admissible at the discrimination points, at least in the initial part of the procedure where a few discrimination points are available.

The initial part of the procedure is of course a key point especially in those cases where no phase measurement is available at all (whereas two phase measurements have been used above). In fact, different combinations of solutions may be admissible along the considered diameter-ring couple. In such a case, as in the crosswords' solution logics, all the different compatible solutions should be retained in the successive steps until the discrimination rules at the subsequent intersection/discrimination points will hopefully allow dropping all (but one) of them.

*V.C. Improvements/1: exploitation of field properties*

The drawbacks listed in the previous Subsection can be tackled as follows:

i. As a first obvious strategy, it is useful starting from a spectral ring having a small radius. In fact, this will reduce the order of the corresponding polynomial, allowing to deal with a reduced number of zeroes and a few field combinations to be checked. In fact, if $H$ is the index to be used in (17), where $H$ is proportional to $k'$, the corresponding square amplitude distributions have $4H$ possible zeroes, paired in couples of the kind $z_i$, $1/z_i^*$ [33]. Then, without using any peculiar property of the spectrum, one should explore a total of $2^{2H}$ possible field behaviors (as any of the $2H$ 'basic' zeroes for the field representation can be both inside or outside the unitary circle in the complex plane). Hence, one as to keep $H$ as small as possible in order to properly trigger the procedure through a reduced number of checks. Note that no analogous reduction of combinations strategy is possible in our previous method [29],[32];

ii. The number of combinations to be checked can be considerably lowered by taking advantage of the properties discussed in Section III.B. In fact, by considering two nearby rings such that $H$ has to be increased by one when moving from the smaller to the larger one, the two new zeroes must necessarily be located one inside and one outside the unitary circle, respectively. Such a circumstance implies that, amongst all the $2^{2H}$ possibilities above (and supposing for the time being that no zero is present in the spectrum within the disk having the radius $k'$ at hand), the ones actually admissible are those where the inherent $2H$ zeroes are located half inside and half outside the unitary circle of the complex plane. Therefore, the number of combinations to be checked considerably reduces from $2^{2H}$ to the binomial coefficient of $\binom{2H}{H}$, thus allowing a second considerable advantage over [29],[32]. Such a property can be exploited in all cases where no zero is present in the pattern over the disk of radius $k'$. However, once one or more zeroes are instead present within the above disk, one can focus the attention on the ring where transition occurs and 'track' the transition of the zeroes from inside to outside and vice versa, so that the number of situations to be checked is still much lower than $2^{2H}$;

iii. Interestingly, the possibilities of enhancing the PR procedure are not over yet. As a matter of fact, the idea of 'zeroes tracking' also can allow for a further reduction of the complexity.

To explain such an idea, let us consider a ring where just $H = 1$ is needed (which happens for a sufficiently small $k'$). In this case, four zeroes are present in the square-amplitude distribution polynomial representation, say $z_1$, $1/z_1^*$, $z_2$, $1/z_2^{*}$[11].

On the basis of considerations above, the number of possible corresponding field behaviors on such a ring is given by the binomial coefficients described above and is equal to 2. In particular, they are associated to the two pairs of zeroes $\{z_1, 1/z_2^*\}$ and (its reciprocal and conjugate) $\{1/z_1^*, z_2\}$.

---

[11] Without any loss of generality, we assume that $z_i$ is within the unitary circle while $1/z_i^*$ is outside it.

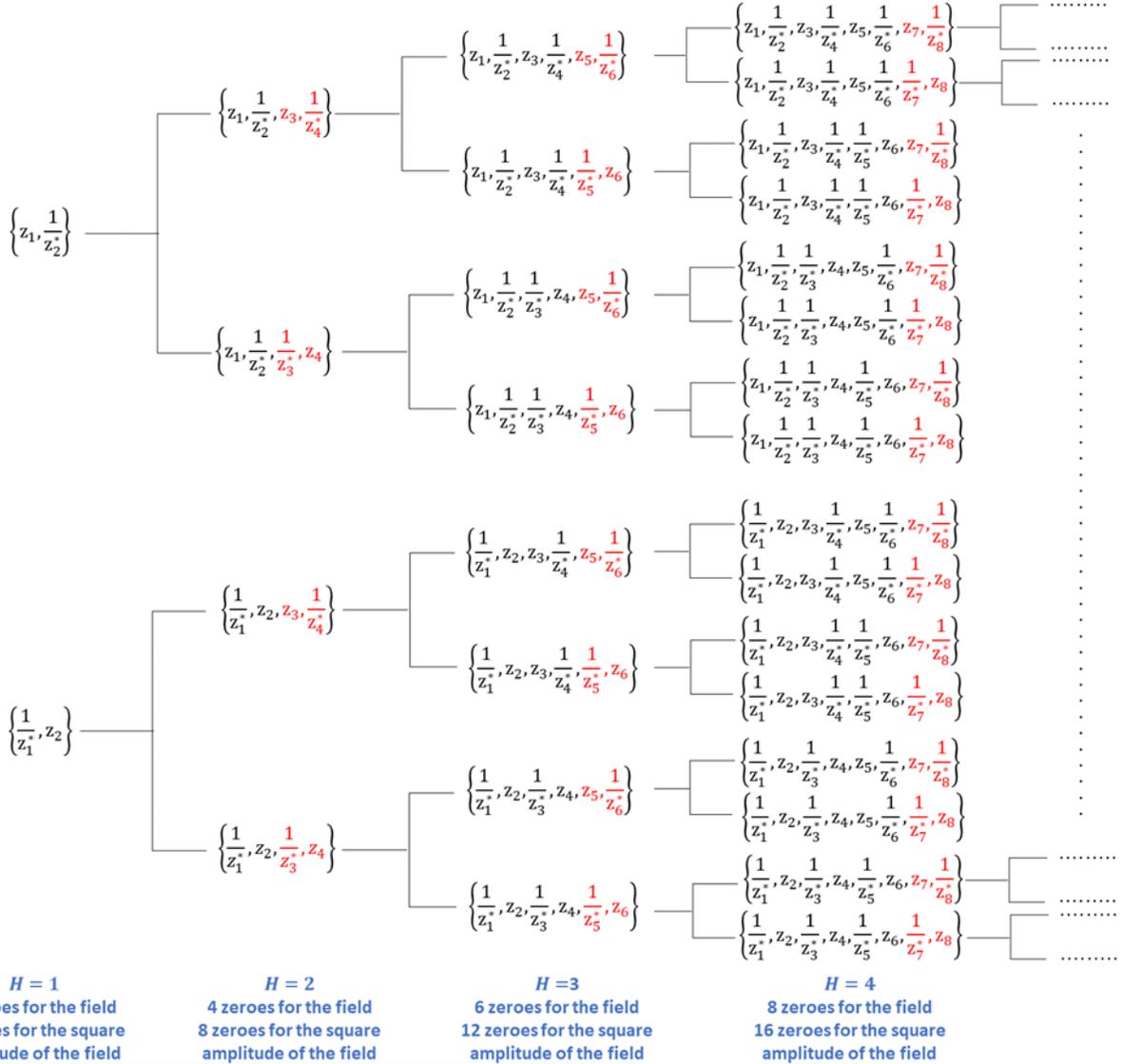

Fig. 2. Bifurcation tree of location of the zeroes when increasing the radius of the ring according to contents of Section V.C. See also footnote 11.

Then, suppose to consider the next ring and that, along it, the field requires $H = 2$. In this case, four additional zeroes are present in the square-amplitude distribution beyond the initial ones, which have gently moved from their original positions in the complex plane. Let us denote by $z_3, 1/z_3^*$, $z_4, 1/z_4^*$ these additional zeroes. Again, by virtue of the field properties, only the couples $\{z_3, 1/z_4^*\}$ and $\{1/z_3^*, z_4\}$ are actually admissible. It follows that each of the initial two solutions 'bifurcates' in two new solutions.

Notably, the same kind of reasoning applies when further increasing the radius of the considered ring. Consequently, by progressively enlarging the rings, the number of potential solutions to be explored follow a rule of the kind $2^H$, which represents a further relevant reduction with respect to the one in the subsection above. An illustrative representation of the bifurcation tree for increasing $H$ values is shown in Fig. 2;

iv. The last (but not least) reduction in the computational burden of the proposed solution procedure can be achieved by pruning the set of possible field behaviors along rings by means of a congruence check along the initially considered diameter (or diameters, at the later stages). In fact, supposing one is able to track the evolution of the zeroes (which is implicitly used in Fig. 2) many of the branches of the tree of the overall possibilities can be simply withdrawn, thus possibly allowing the tracking of a single admissible solution.

*V.D. Improvements/2: a faster exploitation of partial results by means of the Step 2 of the proposed procedure*

Once an adequate number of rings and diameters of the far field have been retrieved through the procedure detailed in Section IV.A, the scenario at hand will appear as depicted in

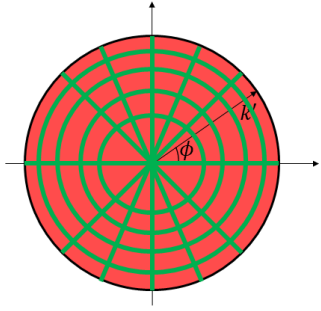

Fig. 3. Pictorial view of the field in the spectral domain after *Step 1* of the proposed procedure: along the green lines the complex field has already been retrieved, while in the red zone it has yet to be recovered.

Fig. 3. In this figure, the retrieved field is highlighted in green while the red part indicates those points where just the amplitude of the field is known. Hence, to run out the PR procedure, we need to recover those points.

Obviously, one can continue to consider additional rings and diameters. In such a way, polynomials of increasing order have to be dealt with when considering larger and larger rings. As an alternative strategy, advantage can be taken from [22],[37]. In these papers it is in fact shown that once a sufficiently large ratio amongst the number of independent square amplitude data and independent number of unknowns is available, false solutions (corresponding to local minima of suitable cost functionals) can be avoided, and PR problems safely solved by local optimization techniques.

Herein, the recovered field samples can be seen as constraints lowering the number of independent unknowns [37], so that the strategies (and procedures) explored and discussed therein can be conveniently applied. In so doing, we take advantage from the minimally redundant field representation $(16a)$ and enforce the already known field behavior by means of penalty terms rather than by strict equality constraints.

Coming to detail, let us suppose splitting the spectral domain samples in two sub-sets: $\hat{\Pi} \equiv (\hat{k}, \hat{\phi})$ where the complex field values have already been retrieved, $\tilde{\Pi} \equiv (\tilde{k}, \tilde{\phi})$ where just amplitude data are known (i.e., $M^2_{\tilde{k},\tilde{\phi}}$). The PR completion is then performed by solving the following optimization problem:

$$\min_{\boldsymbol{b}} \psi(\boldsymbol{b}) = w_1^2 \psi_1(\boldsymbol{b}) + w_2^2 \psi_2(\boldsymbol{b}) \quad (22)$$

where $\boldsymbol{b}$ is the vector containing the representation coefficients $b_{\ell,n}$, while $w_1^2$ and $w_2^2$ are positive constants properly weighting the two functionals $\psi_1$ and $\psi_2$. These latter, following [37], are defined as:

$$\psi_1(\boldsymbol{b}) = \left\| \frac{|F_{\tilde{k},\tilde{\phi}}|^2 - M^2_{\tilde{k},\tilde{\phi}}}{M_{\tilde{k},\tilde{\phi}}} \right\|^2_{\tilde{\Pi}}$$

$$= \sum_{\tilde{k}} \sum_{\tilde{\phi}} \frac{\left[|F_{\tilde{k},\tilde{\phi}}|^2 - M^2_{\tilde{k},\tilde{\phi}}\right]^2}{M^2_{\tilde{k},\tilde{\phi}}} \quad (23a)$$

$$\psi_2(\boldsymbol{b}) = \left\| F_{\hat{k},\hat{\phi}} - T_{\hat{k},\hat{\phi}} \right\|^2_{\hat{\Pi}} = \sum_{\hat{k}} \sum_{\hat{\phi}} |F_{\hat{k},\hat{\phi}} - T_{\hat{k},\hat{\phi}}|^2 \quad (23b)$$

where $T_{\hat{k},\hat{\phi}}$, is the complex field as retrieved in *Step 1*.

Accordingly, the optimization problem (22) aims at determining coefficients $b_{\ell,n}$ such that the amplitude data, i.e., $M^2_{\tilde{k},\tilde{\phi}}$, are fitted through $(23a)$ while $T_{\hat{k},\hat{\phi}}$, is fitted through $(23b)$. Note that while functional $(23b)$ is quadratic with respect to unknowns, functional $(23a)$ is indeed quartic.

To perform the optimization, a gradient-based minimization scheme is adopted. As known, for a given starting guess this kind of deterministic solution algorithm converges to the closest local minimum of the cost functional [53]. However, thanks to *Step 1* and the presence of the corresponding positive definite quadratic functional $(23b)$, one can avoid the occurrence of minima other than the global one.

*V.E. Aperture Field Retrieval*

While the usual PR problem can be considered accomplished at the end of the steps above, the diagnostics problem we are interested in ends with the retrieval of the aperture field distribution since it is the one which actually conveys information about reflector shape deformations.

By virtue of the choice we have performed on the field representation, such a last step can be performed in a simple fashion. In fact, once the complex field is identified, $b_{\ell,n}$ coefficients are also known. Then, we can use (15) in order to determine the aperture field, where $\alpha_{\ell,n} = b_{\ell,n}/\sigma_{\ell,n}$. An even simpler solution, which can be safely applied whenever the spectrum is negligible in its invisible part (which is usually the case with reflector antennas) amounts to perform an inverse 2-D Fourier Transform.

## VI. NUMERICAL EXAMPLES

The aim of this Section is to assess feasibility and effectiveness of the proposed approach in the diagnosis of surface deformations on a reflector antenna. To this end, following [13], we have examined two different cases, namely the case where only the phase of the aperture field is affected and the case where both the amplitude and the phase are affected.

As a reference scenario, we considered a continuous aperture field with a circular support whose nominal expression, corresponding to an undistorted reflector and an out of focus feed, is the following:

$$f(\rho', \phi') = |f|e^{j\varphi_f} \quad (24a)$$

$$|f| = \frac{4FL}{4FL^2 + {\rho'}^2} \quad (24b)$$

$$\varphi_f = \beta \left[ 2FL + \frac{\Delta z}{\rho'}\left(\frac{4FL^2 - {\rho'}^2}{4FL}\right) \right] \quad (24c)$$

wherein $FL$ is the focal length.

In all the experiments, in order to quantitatively appraise the accuracy of the obtained results, we introduced a normalized mean square error (NMSE) metric for both the radiated field (i.e., $NMSE_{rf}$) and the aperture field (i.e., $NMSE_{af}$) as it follows:

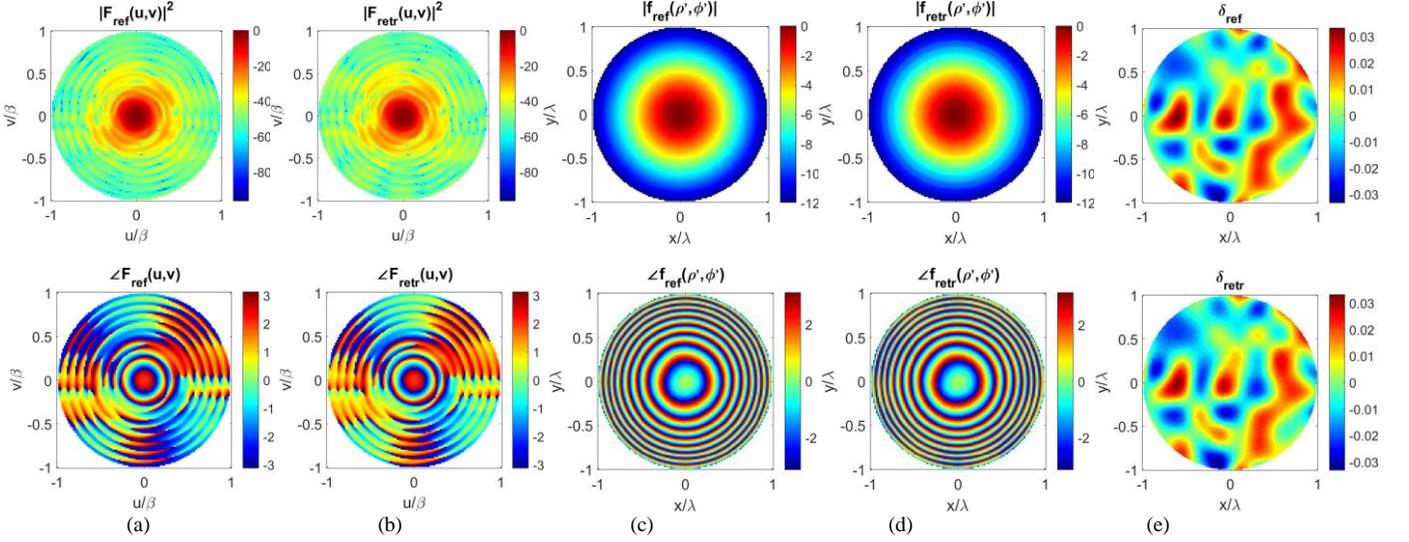

Fig. 4. First numerical example (with $u = k'\cos\phi$, $v = k'\sin\phi$). From top to bottom: amplitude and phase of the reference field (a); amplitude and phase of the retrieved field (b); amplitude and phase of the reference continuous aperture source (c); amplitude and phase of the retrieved continuous aperture source (d); reference and retrieved surface deformation $\delta(\lambda)$ (e).

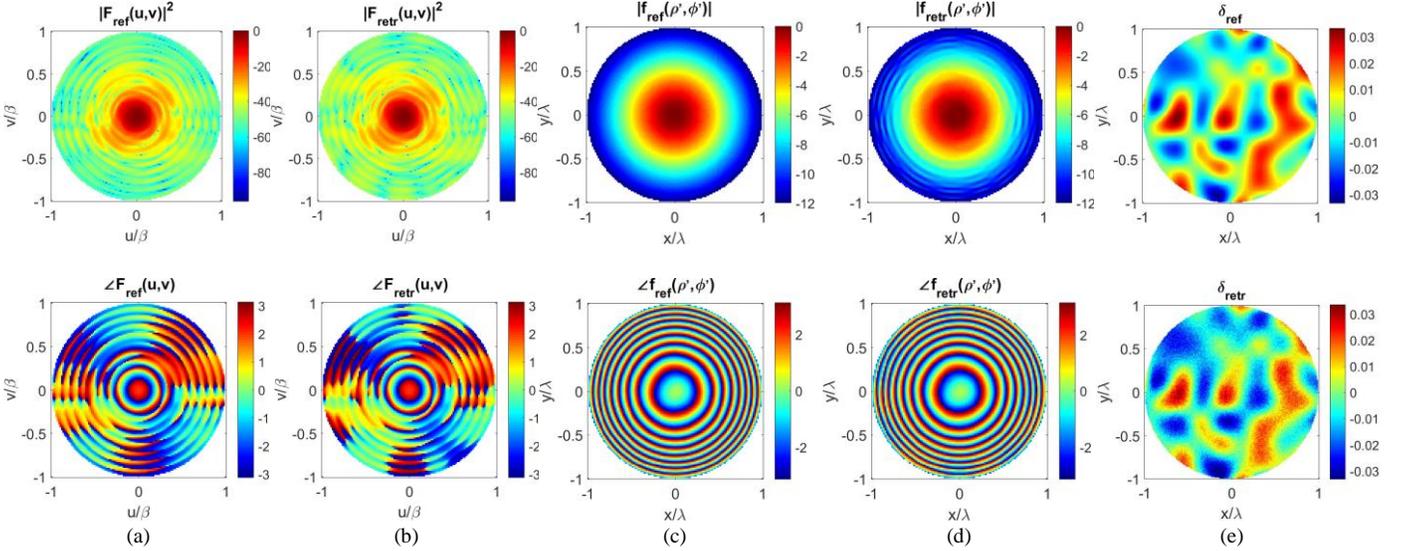

Fig. 5. Second numerical example (with $u = k'\cos\phi$, $v = k'\sin\phi$). From top to bottom: amplitude and phase of the reference field (a); amplitude and phase of the retrieved field (b); amplitude and phase of the reference continuous aperture source (c); amplitude and phase of the retrieved continuous aperture source (d); reference and retrieved surface deformation $\delta(\lambda)$ (e).

$$NMSE_{rf} = \frac{\|F^{actual}(k',\phi) - F^{recovered}(k',\phi)\|^2}{\|F^{actual}(k',\phi)\|^2} \quad (25a)$$

$$NMSE_{af} = \frac{\|f^{actual}(\rho',\phi') - f^{recovered}(\rho',\phi')\|^2}{\|f^{actual}(\rho',\phi')\|^2} \quad (25b)$$

$$f(\rho',\phi') = |f|e^{j(\varphi_f+\Delta)} \quad (26)$$

with:

$$\Delta = \frac{8FL^2\beta}{4FL^2 + \rho'^2}\delta \quad (27)$$

Relationships (26),(27), relating the aperture phase with the surface deformation, have been derived in [2]. As in [13], we set $\delta$ randomly smooth in the range $\left[-\frac{\lambda}{30}, \frac{\lambda}{30}\right]$ [see Fig. 4(e)]. The radius of the source has been set as $10\lambda$, while $FL = 3\lambda$, $\frac{\Delta z}{\rho'} = 0.5$ and the amplitude $|f|$ has been scaled, again in the

In the first example, besides the non-trivial behavior induced by (24c), we also consider, as in [13], a surface deformation $\delta(\rho',\phi')$ on the reflector corresponding to:

same way as in [13], in such a way to exhibit a taper of 12dB.

*Step* 1 of the PR procedure provided the recovery of 1-D complex fields in correspondence of significant diameters and rings of the data matrix. In particular, we have chosen 16 equally spaced rings and 8 diameters ($\phi = [\,0°, 22°, 45°, 68°, 90°, 112°, 135°, 158°]$). In order to initialize the procedure, we have used two complex measurements, which is anyway a very low price as compared to the a-priori information or diversity conditions required in other strategies. As already explained (see Section III.B), these measurements are anyway unessential in the overall logics of the procedure. Also note what we are actually using is the knowledge of the phase shift amongst the two points, so that, as a further alternative, one can also eventually proceed by a trial-and-error procedure on such a value.

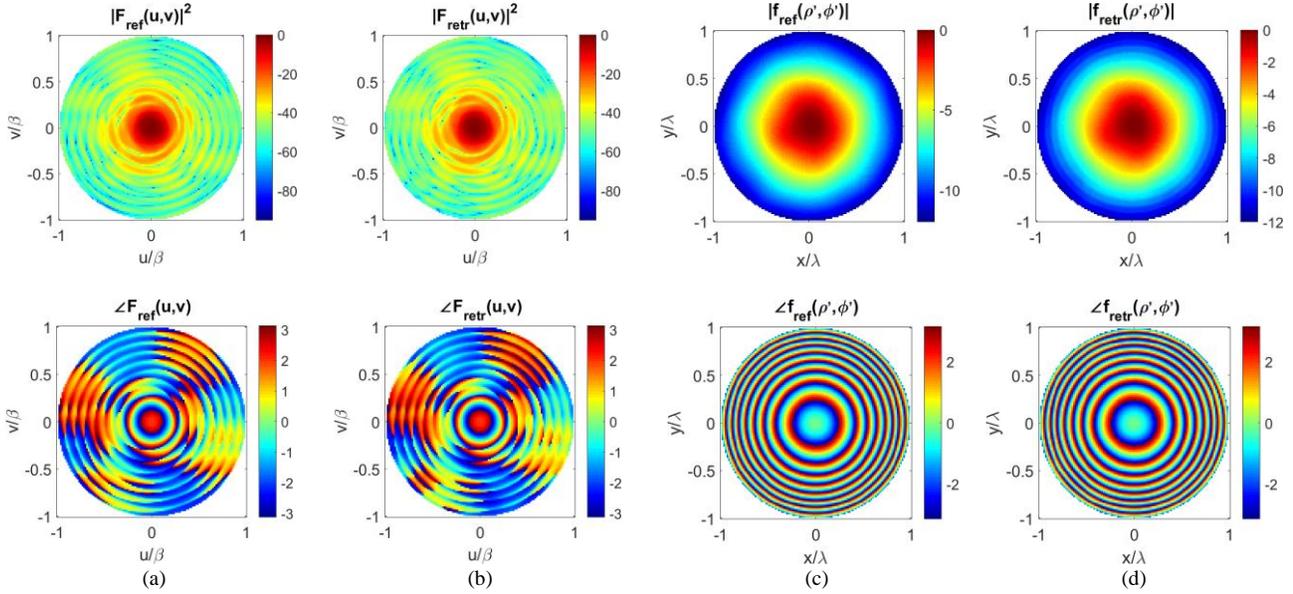

Fig. 6. Third numerical example (with $u = k'\cos\phi$, $v = k'\sin\phi$). From top to bottom: amplitude and phase of the reference field (a); amplitude and phase of the retrieved field (b); amplitude and phase of the reference continuous aperture source (c); amplitude and phase of the retrieved continuous aperture source (d).

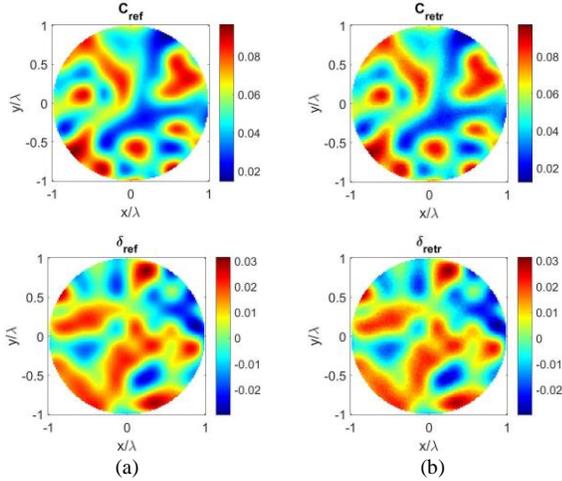

Fig. 7. Third numerical example. From top to bottom: amplitude $C(\lambda)$ and phase $\delta(\lambda)$ of the reference surface deformation (a); amplitude $C(\lambda)$ and phase $\delta(\lambda)$ of the retrieved surface deformation.

Thereafter, the procedure for the PR completion has been undertaken. In particular, we have solved the fitting problem (22), where the weighting parameters $w_1^2$ and $w_2^2$ have been set equal to the energy of the pertaining data. A comparison between reference and reconstructed results is given in Fig. 4 in terms of fields [subplots (a) and (b)], sources [subplot (c) and (d)], and surface deformation [subplot (e)], respectively. As it can be seen, a satisfactory PR solution has been achieved with a $NMSE_{rf} = 4.63 \cdot 10^{-4}$ and $NMSE_{af} = 6.04 \cdot 10^{-4}$. By exploiting a PC equipped with an Intel i7-6700HQ CPU and 16GB RAM, the numerical reconstruction required roughly 80 minutes. As it can be seen, the approach is able to retrieve not only the far field phase, but also the aperture source amplitude and phase [subplot (d)], including the term related to the reflector deformation [subplot (e)].

In the second example, the proposed PR strategy has been successfully tested in case of data corrupted by white gaussian noise with a given signal-to-noise-ratio (SNR). In particular, we used SNR=25dB. Also, the far field has been assumed to be known in the same two points of the spectral plane as before in order to trigger the recovery procedure. The achieved results, corresponding to $NMSE_{rf} = 3.54 \cdot 10^{-3}$ and $NMSE_{af} = 6.36 \cdot 10^{-3}$, are shown in Fig. 5, and confirm the effectiveness of the proposed PR approach. The computational time was approximately equal to the one of the previous noiseless case. Again, one is able to come to a fully satisfactory PR of the radiated field [subplot (b)] as well as of the overall aperture source.

As a second assessment scenario, we simulated a distortion on both the amplitude and the phase of the radiated field. In particular, in the third numerical example, we assumed that the surface deformation $\delta(\rho', \phi')$ [13] causes a phase distortion on $f(\rho', \phi')$, together with an increase $C(\rho', \phi')$ [13] in the amplitude, i.e.:

$$f(\rho', \phi') = |f|(1 + C)e^{j(\varphi_f + \Delta)} \quad (28)$$

According to [13], we set $C$ randomly smooth such that $C < 0.1$ [see Fig. 7(a)]. The radius of the source has again been set as $10\lambda$, while $FL = 3\lambda$ and, also in this case, the amplitude $|f|$ has been scaled in such a way to exhibit a taper of 12dB.

The field pattern has again been assumed completely known in two points of the spectral plane in order to trigger the procedure. Moreover, we have retrieved 19 rings passing through points with high intensity amplitude, and the same 8 diameters of the previous scenario. After that, we have solved the fitting problem (22). A comparison between reference and reconstructed results is given in Fig. 6 in terms of fields [subplots (a) and (b)], and sources [subplot (c) and (d)], respectively, while in Fig. 7 in terms of surface deformation. As it can be seen, the proposed method led again to a fully satisfactory recovery, which is testified by the very low $NMSE_{rf} = 3.26 \cdot 10^{-4}$ and $NMSE_{af} = 5.94 \cdot 10^{-4}$. This numerical example required a computational time of around 90 minutes.

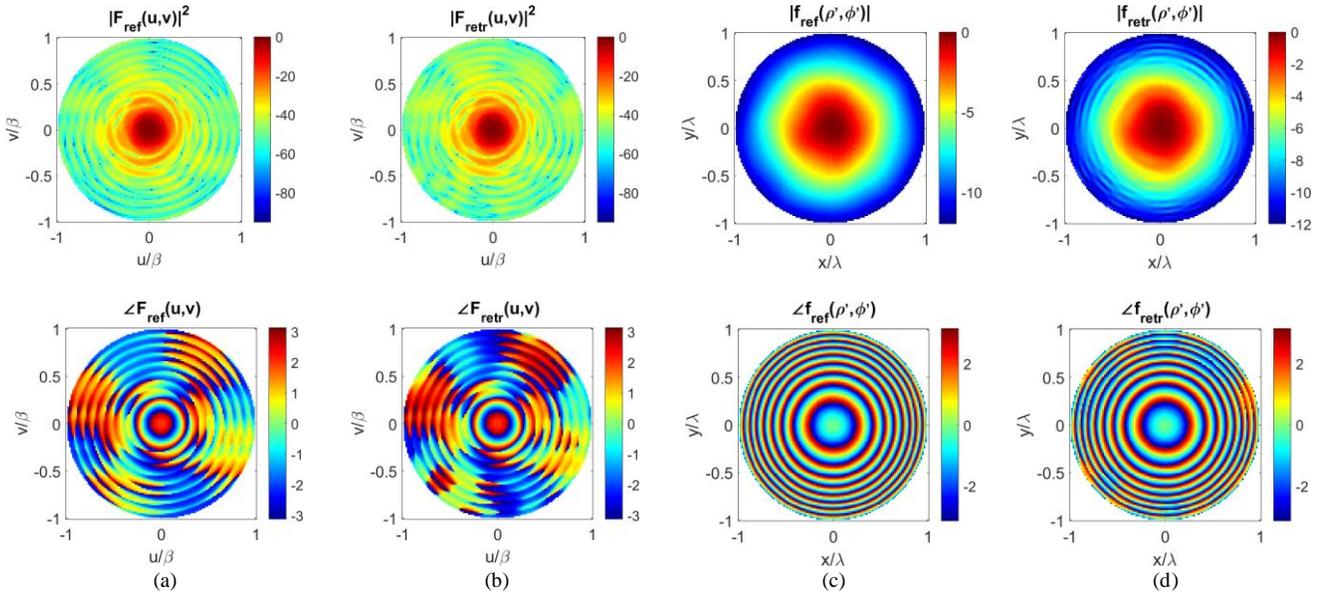

Fig. 8. Fourth numerical example (with $u = k'\cos\phi$, $v = k'\sin\phi$). From top to bottom: amplitude and phase of the reference field (a); amplitude and phase of the retrieved field (b); amplitude and phase of the reference continuous aperture source (c); amplitude and phase of the retrieved continuous aperture source (d).

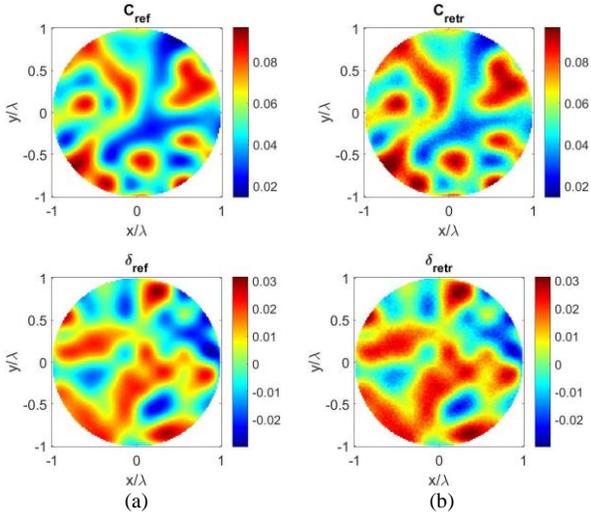

Fig. 9. Fourth numerical example. From top to bottom: amplitude $C(\lambda)$ and phase $\delta(\lambda)$ of the reference surface deformation (a); amplitude $C(\lambda)$ and phase $\delta(\lambda)$ of the retrieved surface deformation.

In the fourth and last test case, we used the same source of the previous example but enhanced the PR difficulty by corrupting each measured amplitude with SNR=25dB. In this example, the knowledge of amplitude and phase in four points of the spectral plane has been assumed in order to trigger the procedure (two along the horizontal diameter $\phi = 0$ and two along the vertical diameter $\phi = 90°$). A comparison between reference and reconstructed results is given in Fig. 8 in terms of fields [subplots (a) and (b)] and sources [subplot (c) and (d)], respectively, while in Fig. 9 in terms of surface deformation. As in all the previous test cases, a satisfactory PR solution has been achieved as testified by $NMSE_{rf} = 4.93 \cdot 10^{-3}$ and $NMSE_{af} = 7.03 \cdot 10^{-3}$. As in the second test case, no difference in terms of computational time has been experienced with respect to the previous noiseless case.

## VII. CONCLUSIONS

A new general approach to the canonical problem of reconstructing 2-D fields starting from amplitude measurements has been proposed and discussed.

The proposed technique, which also aims to recover the corresponding aperture sources, jointly exploits a number of theoretical and methodological tools including recent expansions based on OAM modes and SVD, the overlooked spectral factorization method, and a crosswords-inspired processing. By complementing all the above with analysis and exploitation of the zeroes of the trigonometric polynomials expressing the field and its power pattern along diameters and rings, the proposed procedure brings decisive advantages with respect to the state-of-the-art approaches in terms of actual number of required measurements. In fact, differently from essentially all the available phase retrieval procedures, the presented one just requires a single measurement set (plus the knowledge of the support of the source and some minimal a-priori information), with relevant advantages in terms of cost, reliability, and effectiveness.

The actual capability to retrieve the complex aperture field distribution has been assessed in the detection of shape deformations on reflector antennas. It is expected that a full exploitation of all the discussed possible improvements as well as the hybridization of the proposed approach and tools with other methods, will further boost the effectiveness and interest of the proposed technique.

APPENDIX A

PROPERTIES OF FIELD ZEROES BY MOVING ALONG RINGS WITH INCREASING RADIUS

Let us consider the polynomial representation (17) for the field. For $k' = 0$, it particularizes as:

$$F(0, \phi) = C_0 \qquad (A1)$$

where we can assume $C_0 = 1$ for the sake of simplicity. Then, let us suppose to move to a ring in the close proximity of the previous one, that is $k' = \varepsilon$ ($\varepsilon$ denoting a real and positive small number). In this case, we expect that:

$$F(\varepsilon, \phi) = \sum_{\ell=-1}^{1} C_\ell\, z^\ell \qquad (A2)$$

with $z = e^{j\phi}$, i.e., the field representation admits two zeroes, $z_1$ and $z_2$. Accordingly, (A2) can be re-written as follows:

$$F(\varepsilon, \phi) = C_0 (z - z_1)\left(\frac{1}{z} - \frac{1}{z_2}\right) \qquad (A3)$$

For $\varepsilon$ sufficiently small, the field behavior should keep almost unchanged, i.e.:

$$(z - z_1)\left(\frac{1}{z} - \frac{1}{z_2}\right) \approx C_0 = 1$$
$$1 - \frac{1}{z_2}e^{j\phi} - z_1 e^{-j\phi} + \frac{z_1}{z_2} \approx 1 \qquad (A4)$$

which holds true if $z_1 \to 0$ and $|z_2| \to \infty$.

By supposing that zeroes gently vary by moving along rings having an increased radius, and that no zero is present in the pattern over the disk of radius $k'$ (otherwise the two zeroes degenerate to a zero having multiplicity two on the unitary circle of the complex plane), the same arguments can be applied by gradually increasing the radius (and the corresponding number $H$ of harmonics).

As a consequence, result (A4) allows asserting that the couple of zeroes coming into play by increasing the radius of the ring emerge from zero and infinity, respectively. Hence, one of them is located inside the unitary circle of the complex plane, while the other one is located outside it.